\documentclass[a4paper,fleqn,usenatbib]{mnras}
\usepackage{amssymb,amsmath,graphicx,psfrag,mathtools}
\usepackage[T1]{fontenc} 
\usepackage{ae,aecompl} 
\usepackage{url}
\title[Are Fast Radio Bursts Made By Neutron Stars?]{Are Fast Radio Bursts
Made By Neutron Stars?}
\author[J. I. Katz]{
J. I. Katz,$^{1}$\thanks{E-mail katz@wuphys.wustl.edu} 
\\
$^{1}$Department of Physics and McDonnell Center for the Space Sciences,
Washington University, St. Louis, Mo. 63130 USA 
}
\date{Accepted XXX.  Received YYY; in original form ZZZ} 
\pubyear{2019} 
\date{\today}
\begin{document} 
\label{firstpage} 
\pagerange{\pageref{firstpage}--\pageref{lastpage}} 
\maketitle 
\begin{abstract}
	Popular models of repeating Fast Radio Bursts (and perhaps of all
	Fast Radio Bursts) involve neutron stars because of their high
	rotational or magnetostatic energy densities.  These models take one
	of two forms: giant but rare pulsar-like pulses like those of
	Rotating RAdio Transients, and outbursts like those of Soft Gamma
	Repeaters.  Here I collate the evidence, recently strengthened,
	against these models, including the absence of Galactic micro-FRB,
	and attribute the 16 day periodicity of FRB 180916.J0158+65 to the
	precession of a jet produced by a massive black hole's accretion
	disc.
\end{abstract}
\begin{keywords} 
	radio continuum: transients, stars: neutron, (transients:) fast
	radio bursts, stars: black holes 
\end{keywords} 
\section{Introduction}
The sources and mechanisms of Fast Radio Bursts (FRB) are one of the most
prominent mysteries of modern astronomy.  Most models involve neutron stars
to take advantage of their deep gravitational potential wells, the great
magnetostatic and rotational energies of some neutron stars and their other
known transient emissions.  Pulsar-like models provide a natural analogy to
the coherent emission of FRB.  Magnetostatic energy (``magnetar'') models of
Soft Gamma Repeaters (SGR) readily provide the energies of FRB, as may
neutron star accretional models involving the release of gravitational
energy.  { Although many models of pulsars and magnetars have been
developed, none of them} led to a prediction of FRB, {as might have been
expected were FRB their natural consequence.  These models} require very
large extrapolations, quantitative (in energy) or qualitative (in the type
of emission) { to account for FRB, which suggests re-examination of the
assumption that neutron stars are responsible}.

Neutron star models have difficulty explaining repeating FRB because the
well-studied repeating FRB 121102 is not periodic \citep{Z18}.  Magnetic
fields are essential to pulsar and ``magnetar'' SGR models {of FRB.
Magnetic fields vary with direction from their source, as will any radiation
related to the field.  Unless a rotationally aligned dipole, rotation sweeps
the {angular distribution of radiation emitted near the neutron star}
across the observer, leading to an observed periodic modulation or
recurrence at integral multiples of an underlying period, whatever the
radiation mechanism.  Examples include pulsars, Rotating RAdio Transients
(RRAT), SGR and Anomalous X-Ray Pulsars (AXP, the quiescent phase of SGR).}

{Radiation energized by a neutron star's rotationally swept fields or
particles may be emitted far away, perhaps in a wind nebula or supernova
remnant.  Brief bursts, like FRB, emitted by interaction with distant small
structures would also be rotationally modulated, although continuous
emission need not be.}

These difficulties arise in any neutron star model that involves a magnetic
field: pulsar-like, SGR-like and accretional models in which a magnetic
field channels accretion.  They apply also to apparently non-repeating FRB
if they are, as suggested but unproven, repeaters whose repetitions have not
been observed because of their infrequency \citep{J19} or insufficient
observational coverage.

FRB were reviewed by \citet{K16a,K18a,CC19,PHL19}; \citet{P18} provides a
complete and updated catalogue of proposed models.  The argument of the
preceding paragraphs is not universally accepted, and neutron star models
remain popular.  It is a strong argument against pulsar-like models, whose
rotation implies bursts separated by integer multiples of a rotation period.
It is a somewhat weaker argument against SGR-like models in which it only
implies periodic modulation of the observed strengths and frequencies of
bursts.  Although AXP are periodically modulated and longer SGR outbursts
show periodic substructure, rotational modulation of the timing of detected
brief SGR outbursts is not evident.

The purpose of this note is to synthesize the theoretical and observational
arguments against any neutron star origin of FRB.  I pay particular
attention to the new upper bounds on MeV gamma-ray emission of two repeating
FRB found by \citet{C19} that provide additional evidence against SGR-like
models.
\section{Pulsar-like models of FRB}
In these models FRB are produced by the same mechanisms as radio pulsars,
but with much higher energies and with most pulses nulled; they would be
more energetic analogues of RRAT.  Such models imply pulse intervals that
are integer multiples of a neutron star's rotation period.  This appears to
be inconsistent both with older data \citep{H17,S17} and with a series of 93
bursts observed in one five-hour observing run of the repeater FRB 121102
\citep{Z18}.

Such a run is short enough that plausible period derivatives do not break
the requirement that burst separations be integer multiples of a single
period.  Timing of bursts separated by gaps longer than a few hours cannot
constrain short (ms) periods because plausible period derivatives make the
cycle count ambiguous, although the different short periods derived from
different runs must be consistent with plausible spindown rates.  Intervals
between bursts in widely separated runs can constrain longer periods, but
these have been excluded for FRB 121102 on the basis of the multiple
intervals observed in a single run; see discussions in \citet{K18b,K19}.  

Energetics are an additional problem for pulsar-like models.  The usual
assumption that pulsars have no energy reservoir between their rotational
energy, tapped at the rate of dipole radiation, and a relativistic wind and
radiation field, implies extreme values of both magnetic dipole moment and
rotation rate in order to explain FRB powers $\sim 10^{43}$ ergs/s.  This
combination may be impossible, and would imply very short lifetimes
\citep{K16a,K18a}.

There are two possible loopholes to the energetic argument: If FRB are
narrowly collimated \citep{K17a,K17b} their power requirements would be
correspondingly relaxed.  If pulsar magnetospheres contain an intermediate
energy reservoir, such as might be provided by transitions \citep{K17c}
between the magnetospheric states of intermittent pulsars \citep{K06} whose
spindown rates differ by tens of percent and pulse powers by orders of
magnitude, their dipole moment and spin rate would be essentially
unconstrained.  Both these loopholes are speculative, and there is no
evident path to closing them.
\section{SGR-like models of FRB}
SGR-like models are attractive because of their abundant energy; the giant
outburst of SGR1806-20 on December 27, 2004 released about $10^{47}$ ergs in
about 0.1 s \citep{H05,P05}.  This is about seven orders of magnitude
greater than energies inferred for FRB \citep{T13}, and the ratio is even
larger if FRB are collimated, as is plausible for coherent radiation by
relativistic particles.  In addition, SGR have sub-ms rise times (see
discussion in \citet{K16b}), consistent with the $\sim$ ms durations of FRB
and shorter than any other known astronomical process other than pulsar
pulses and their substructure.
\subsection{Theoretical difficulties}
{ Here I consider issues that arise if FRB are produced by relativistic
electrons near the surfaces of neutron stars around the peaks of SGR
outbursts.}  SGR appear to be thermalized sources with approximately
black-body spectra at temperatures of tens or hundreds of keV, while FRB are
produced by coherent non-thermal processes with brightness temperatures as
high as $\sim 10^{35}$ K.  In general, uncollimated radiation intensities
$\gtrsim 10^{29}$ ergs/(cm$^2$-s), about 10$^{-6}$ of the intensity of SGR
1806-20 at a neutron star radius, rapidly thermalize into equilibrium
photon-pair plasma \citep{K96}.  The spectral data on SGR are averaged over
their $\sim 0.1$ s durations and may not constrain their spectra during
their sub-ms rise { or at other times when their luminosity is low, so these
issues may not arise if FRB are emitted when the SGR luminosity is below its
peak}.

The radiation environment of a SGR { during the peak of its outburst} is
hostile to relativistic particles, such as required {in many models} to
radiate a FRB.  Particles radiating curvature radiation\footnote{An
alternative hypothesis, in which FRB are analogous to Type III Solar radio
bursts, suffers from the problem that these are produced in plasma whose
density is at least 1/4 of the critical density at the frequency of
emission.  As a result the dispersion index will not be close to 2, in
conflict with observation, unless the emission region has a very small scale
height and its contribution to the dispersion is negligible.} at a frequency
$\nu$ in a magnetic field with radius of curvature $R$ have Lorentz factors
\begin{equation}
	\gamma \sim \left({\nu R \over c}\right)^{1/3} \sim 50,
\end{equation}
where we have taken $\nu \sim 1$ GHz and $R \sim 10^6$ cm, appropriate to
neutron star models of FRB.  A relativistic electron of energy $E = \gamma
m_e c^2$ moving through a thermal uncollimated radiation field of energy
density $\cal E$ suffers an energy loss by Compton scattering
\begin{equation}
	{dE \over d\ell} \sim \gamma^2 \sigma {\cal E},
\end{equation}
where $\ell$ measures its path and $\sigma$ is the Compton energy loss
scattering cross-section (the Klein-Nishina cross-section convolved with the
kinematics of recoil energy loss)
\begin{equation}
	\sigma \sim \left({e^2 \over m_e c^2}\right)^2 {\ln{(E_p/m_e c^2)}
	\over E_p/m_e c^2} \sim \left({e^2 \over m_e c^2}\right)^2
	{\ln{\gamma} \over \gamma},
\end{equation}
where $E_p$ is the photon energy in the electron's frame.  The final
approximation applies to a photon with $h\nu \sim m_ec^2$ in the star's
frame, a representative value for a black body spectrum characteristic of
the giant outburst of SGR 1806-20, for which $E_p \sim \gamma m_e c^2$.
The energy loss length
\begin{equation}
	\label{losslength}
	\ell \sim {(m_e c^2)^3 \over e^4 {\cal E} \ln{\gamma}} \sim 10^{-7}
	\left({10^{25}\,\text{erg/cm}^3 \over {\cal E}}\right)\,\text{cm}.
\end{equation}
A SGR emitting $P \sim 10^{48}$ erg/s \citep{H05,P05} from the $A \sim
10^{13}$ cm$^2$ surface area of a neutron star has ${\cal E} = 4P/(Ac) \sim
10^{25}$ erg/cm$^3$; energy loss is extremely rapid.

In order to make up this energy loss by acceleration would require an
electric field
\begin{equation}
	E_{el} \sim {\gamma m_e c^2 \over e\ell} \sim 10^{12}\,
	\text{esu/cm}.
\end{equation}
Such a field cannot be realized.  In vacuum it would rapidly lead to
breakdown into a pair gap, as in standard pulsar theory.  In the dense
equilibrium pair plasma (${\cal E} \sim 10^{25}$ erg/cm$^3$, $n_\pm \sim
10^{31}$ cm$^{-3}$) at temperature $k_B T \sim m_e c^2$ required for the
emission of $\sim 10^{35}$ erg/(cm$^2$-s) observed in SGR 1806-20 it would
imply the impossible power density $E_{el} e n_\pm c \sim 10^{43}$
erg/(cm$^3$-s).
\subsection{Observational difficulties}
\citet{Me19} set upper bounds on the rate of FRB at the locations of six
gamma-ray bursts suggested to house ``magnetar'' neutron stars.
The failure of \citet{TKP16} to detect a FRB during a fortuitous observation
of the great outburst of SGR 1806-20 is a strong argument against the
association of FRB with SGR, although collimation of the FRB is a possible
loophole.  The recent results of \citet{C19} make the converse argument
against the association of FRB with SGR: The AGILE X-ray and gamma-ray
satellite viewed two repeating FRB during their outbursts, and no X- or
gamma-rays were observed, with upper limits $\sim 2 \times 10^{46}$ ergs at
the {distance of 149 Mpc of FRB 180916 \citep{M20}}.  This is inconsistent
with an outburst like the great
outburst of SGR 1806-20.  \citet{C19}'s Source 2 at $\sim 300$ Mpc (an upper
bound on its distance implied by its dispersion measure, assuming only
standard cosmology) {is also} likely inconsistent with an event like SGR
1806-20.   This argument cannot be evaded by collimation because the thermal
soft gamma-ray emission of SGR cannot be {strongly} collimated.  {
On the other hand, these observational bounds on the ratios of MeV to radio
fluxes are consistent with those predicted \citep{B19,M19} in some
magnetar/neutron star models of FRB.}

One such giant SGR outburst has been observed in the $\approx$ 50 years
since the launch of the Vela satellites, corresponding to a rate of $\sim$
0.02/year in the Galaxy.  The FRB rate per galaxy is much less than this.
Although we do not know the luminosity function of SGR giant outbursts, 
there {may be} rare outbursts significantly stronger than even the once per
$\sim$ 50 years great outburst of SGR 1806-20.  If FRB are associated with
SGR, the strongest and most observable FRB would {plausibly} be associated
with the most luminous SGR.  Association of the repeating FRB observed by
\citet{C19} with such a super-SGR 1806-20 outburst is empirically excluded.
\section{FRB Sky Distribution}
The distribution of FRB on the sky shows no evidence of a Galactic
contribution.  In contrast, the sky distribution of integrated fluence of
every other extra-Solar System astronomical radiation of stellar origin,
with the sole exception of gamma-ray bursts (GRB), is dominated by the
Galactic disc.  If observations could be extended for the Galactic GRB
recurrence time, it is expected that the Galactic disc would also dominate
the GRB fluence.  This is a consequence of the dominance of the baryonic
mass distribution of the Universe, weighted by the inverse square of
distance, by the Galactic disc.  It would apply to FRB if they are produced
by sources related to stars, provided that even {\it one\/} were present in
the Galaxy.

A FRB (or any event) at a Galactic distance of 10 kpc would be about 117
dB brighter than at cosmological ($z = 1$; luminosity distance of about 7
Gpc) distances.  The far ($\sim 60^\circ$) side-lobe sensitivity of radio
telescopes is typically about 60 dB less than their mean beam sensitivity,
leaving about 57 dB of headroom for detection of Galactic micro-FRB.  They
would be detected, with comparable signal-processing systems, in any
pulse-sensitive observation if their intrinsic strength (some appropriate
combination of radiated power and spectral energy density) were within five
orders of magnitude of those of the observed cosmological FRB.  Most
non-catastrophic transient phenomena (those that do not destroy their
sources) in Nature, such as Solar and stellar flares, earthquakes,
lightning, SGR outbursts and giant pulsar pulses, have a wide range of
strengths, as do FRB \citep{Ku19}, with weak events far more numerous than
strong ones.

If the differential size distribution of FRB is a power law $dN/dE \propto
E^{-\alpha}$ then the rate of FRB detectable at $\sim 10$ kpc would be $\sim
10^{11.7(\alpha-1)}$ times the Galactic rate of FRB strong enough to be
detectable from $z = 1$, or $\sim 10^{11.7\alpha-14.5}$/y.  The exponent
$\alpha$ describes the distribution of strengths of all FRB, including those
from sources that are too weak to be detectable at cosmological distances as
well as weak FRB from sources (like FRB 121102) whose stronger bursts are
detectable at those distances.  Therefore, $\alpha$ is likely to be greater
than the exponent fitted to the distribution of bursts from an individual
source, such as FRB 121102 \citep{G19,W19}.

The absence of detected Galactic micro-FRB implies that the Galaxy contains
{\it no\/} objects that could emit repeating FRB.  It argues against neutron
stars as sources because there are many, with ranges of several orders of
magnitude of magnetic fields, rotation rates and ages, in the Galaxy; if
neutron stars with optimal values of parameters make FRB that can be
detected at $z \sim 1$, neutron stars with less optimal values of parameters
should emit micro-FRB detectable at $\sim 10$ kpc.

This argument does not apply to catastrophic models of FRB (just as it does
not apply to supernov\ae\ or GRB, that are catastrophic) because in such
models there are no micro-FRB (just as there are no micro-SN or micro-GRB).
The FRB rate, like the GRB rate, would be so low that {\it none\/} would
likely have occurred during the period of observations.  Could we integrate
long enough, the FRB fluence, like the GRB fluence, would be dominated by
the Galactic disc.  But we know that repeating FRB cannot be catastrophic.

The choice of a nominal Galactic distance of 10 kpc assumes that the
Galactic FRB rate is not dominated by even closer and weaker but more
frequent or abundant sources.  This holds for the Galactic disc
(independently excluded by the isotropic distribution of FRB) if $\alpha <
3/2$ and for the isotropic immediate ($\lesssim 100$ pc) Solar neighborhood
if $\alpha < 2$.  If these conditions are not met, FRB sources must be rare
enough that there are {\it none\/} within those distances, which has the
same effect as requiring a low $E$ cutoff on $dN/dE$.  This is a weaker
version of the inference that there are no FRB sources within the Galaxy;
they are discrete, their number density is finite and their spatial density
is cut off at the statistically expected mean distance of the nearest one.
\section{FRB 180916.J0158+65}
The recently discovered \citep{CHIME20} $P = 16.35$ day modulation
period of FRB 180916.J0158+65 is much too long to be ascribed to the
rotation of a neutron star, whose known rotation periods are $\sim
10^{-3}$--$10^3$ s.  It might be a binary orbital or superorbital (disc
precession or apsidal advance) period, 10--50 times longer than the orbital
period, in analogy to Her X-1, Cyg X-1 and SS433; it neither requires nor
excludes a neutron star.  The apparent absence of an analogous long period
in FRB 121102 may perhaps be attributed to its comparatively few and
scattered (though longer) observations, in contrast to the approximately 300
observations of FRB 180916.J0158+65 well distributed over a year
\citep{CHIME20}.

If this period is orbital \citep{IZ20,LBG20}, the orbit is circular, the
total mass of the binary is $M$ and the variation of DM around the orbit is
$< \Delta$DM then the characteristic value of the electron density in the
orbit is bounded:
\begin{equation}
	\begin{split}
	n_e &< \Delta\text{DM}\left({4 \pi^2 \over G M P^2}\right)^{1/3}\\
	&\lesssim 1.5 \times 10^5 \text{cm}^{-3} {\Delta\text{DM} \over
	0.1\ \text{pc/cm}^3} \left({M \over M_\odot}\right)^{-1/3}.
	\end{split}
\end{equation}
A corresponding bound on a mass flow rate may be estimated
\begin{equation}
	{\dot M} \sim n_e m_p R^2 v \lesssim 10^{13}
	{\Delta\text{DM} \over 0.1\ \text{pc/cm}^3}
	\left({M \over M_\odot}\right)^{2/3}\ \text{g/s},
\end{equation}
where $R$ is the orbital radius and $v$ a flow speed; this, of course,
assumes a roughly isotropic wind and does not constrain denser flows that
do not intersect our line of sight.  This may be related to the absorption
of burst radiation, but our ignorance of the plasma temperature and $M$,
that may be $\gg M_\odot$, precludes quantification.

The confinement of bursts within about 0.3 of the period, as opposed to a
smoother modulation of their rate, suggests intermittent activity in a
precessing beam produced by black hole accretion \citep{K17b}, in analogy to
the precession of jets in AGN \citep{L90,C04} and SS433 \citep{M82}.  Their
possible relation to FRB is supported by the inference of offset massive
black holes in dwarf galaxies \citep{RCDG20} and the identification of FRB
121102 with a dwarf galaxy \citep{T17} and the offset of FRB 180924
\citep{Ba19} and FRB 190523 \citep{R19} from the centers of their host
galaxies.
\section{Discussion}
Neutron star models of repeating FRB are specious.  Pulsar-like models imply
periodicity that is not observed.  They make energetic demands that are
difficult to meet.  SGR-like models imply periodic modulation that has not
been seen.  More importantly, no FRB was observed in association with a
Galactic SGR and SGR are excluded from association with two extragalactic
FRB.  Repeating FRB require a different explanation.

If apparently non-repeating FRB are actually one-off, catastrophic events
these arguments would not apply to them.  There would need to be two
different FRB mechanisms, one for repeaters and one for non-repeaters; the
latter could involve the birth or death of a neutron star.

The rarity of FRB sources implied by the absence of Galactic micro-FRB
excludes stellar mass black holes as well as neutron stars (unless they are
so narrowly collimated that none in our Galaxy are observable).  A neutron
star model might satisfy the constraint of rarity (but not that of
aperiodicity) by limiting emission to the very youngest and perhaps fastest
rotating or most strongly magnetized stars.  No such loophole exists for
black holes, whose properties (aside from mass and angular momentum if they
are rapidly accreting) do not change with age.  The only known objects rare
enough to meet the criterion of rarity are intermediate-mass or massive
black holes \citep{K19}.

Comparison to FRB 121102 argues against attributing the absence of Galactic
neutron star FRB to short active lifetimes.  FRB 121102 has been active for
seven years, with no apparent sign of decay; a neutron star's activity, even
if decaying, would remain observable at Galactic distances very much longer
than at the distance of FRB 121102 at which it would be $\sim 10^{10}$ times
fainter.  There are likely between 30 and 300 Galactic neutron stars, with a
wide range of parameters (magnetic field, binary companions, spin, {\it
etc.\/}), younger than $10^4$ years; any as energetic as FRB 121102 would be
brighter than it unless its radiated flux decayed faster than the
$-10/\log_{10}{(10^4\,\text{y}/T_{age})} < -3$ power of time, where
$T_{age}$ is the present age of FRB 121102.  If FRB sources are neutron
stars, they must in some way be distinguished from the overwhelming majority
of neutron stars, such as by orientation if FRB emission is narrowly and
stably beamed.

Precession of a beam and the disc that feeds and guides it can be driven by
the Lense-Thirring effect or nonrelativistically by a massive surrounding
disc.  In contrast to models \citep{LBB20,ZL20} based on free precession of
a neutron star that predict a smoothly lengthening precession period as the
star spins down, and binary models \citep{IZ20,LBG20,YZ20} in which orbital
and precession periods are stable, models based on a precessing jet produced
by black hole accretion are consistent with any trend unless the driving
disc is dissipating.  A disc remnant of a disrupted star \citep{NK13} would
gradually dissipate, the torque it exerts would decline, and the resulting
precession period would lengthen.  The observed \citep{CHIME20} maintenance
of phase stability in FRB 180916.J0158+65 to $\Delta \phi \lesssim 1$ radian
over an observation time $t_{obs}$ implies a lower bound on a characteristic
time scale of steady period change $t_{char} \equiv P/|{\dot P}| \gtrsim 2
\pi (t_{obs}/2)^2/(2 P \Delta \phi) \sim 20$ y.

If FRB are produced in accretion disc funnels or jets, analogous phenomena
might be observable in blazars, in which these funnels and jets are directed
to the observer, although their dependence on black hole mass, accretion
rate and other parameters is unknown.
\section*{Acknowledgements}
I thank an AAS Foreign Travel Grant for enabling participation in the 2018
Fast Radio Burst Workshop at the Weizmann Institute, for and at which these
ideas began gestating.

\bsp 
\label{lastpage} 
\end{document}